# Phenomenological Modeling of Memristive Devices

F. Merrikh-Bayat[1], B. Hoskins[1,2], and D.B. Strukov[1*]

[1]Department of Electrical and Computer Engineering, University of California, Santa Barbara, CA 93106

[2]Materials Department, University of California, Santa Barbara, CA 93106

Email: * strukov@ece.ucsb.edu

**Abstract**

We present a computationally inexpensive yet accurate phenomenological model of memristive behavior in titanium dioxide devices by fitting experimental data. By design, the model predicts most accurately *I-V* relation at small non-disturbing electrical stresses, which is often the most critical range of operation for circuit modeling. While the choice of fitting functions is motivated by the switching and conduction mechanisms of particular titanium dioxide devices, the proposed modeling methodology is general enough to be applied to different types of memory devices which feature smooth non-abrupt resistance switching.

**Index Terms** – RRAM, memristive behavior, modeling, titanium dioxide devices

## I. Introduction

The recent progress in resistive switching devices [Was09, Lee11, Str11, Tor11, Jam13, Gov13, ITRS13] gives hope for adoption of this technology in various computing applications [Str12b, Yan13] in the near future. The development of such applications, and in particular those utilizing analog properties of memristive devices [Ali12, Ali13], will heavily rely on the availability of accurate predictive device models. Ideally, such models should describe complete *I-V* behavior, e.g. being able to predict current $i(t_0)$ via device at time $t_0$ for an applied voltage bias $v(t_0)$. Because resistive switching devices have memory, the current $i(t_0)$ should also depend on the history of applied voltage bias before time $t_0$, or, equivalently, on the memory state variable vector $\boldsymbol{w}$ at time $t_0$. Such memory state variables represent certain physical parameters, which are changing upon switching the device, e.g. radius and/or length of switching filament [Pic09, Yu11, Gao11, Men12]. A very convenient method for capturing the complete *I-V* behavior is to use a set of two equations describing memristive system [Chu76]. In particular, the change in memory state of a device is described as a function of applied electrical stimulus (e.g. voltage bias) and the current memory state of the device, i.e.

$$\dot{\boldsymbol{w}} = G(v(t), \boldsymbol{w}). \qquad (1)$$

The static equation models current-voltage relation for a particular memory state, i.e.

$$i = F(v(t), \boldsymbol{w})v(t). \qquad (2)$$

While there has been an impressive progress in understanding and modeling switching behavior in memristive devices, the majority of reported models are not suitable for large-scale circuit simulations. For example, some models focus on specific aspects of the memristive behavior, e.g. static equation only [Ber10], or a particular aspect of switching dynamics [Iel11, Str12a, Gao11, Jam11], and hence are incomplete. Others are derived assuming very simple physical models [Bio09, Hur10, Yac13] and therefore could not accurately predict experimental behavior - see also comprehensive reviews of such models in Refs. [Esh12, Kva13]. Alternatively, some models are too computationally intensive, e.g. due to necessity of solving coupled differential equations [Jeo09, Str09a, Gua12a, Lar12, Kim13, Kim14, But13, Mic13, Mic14] or running molecular dynamics simulations [Cha08, Sav11, Pan11]. Several compact (SPICE) models, which are the most suitable for large scale simulations, have been also very recently proposed for valence change [Pic09, Abd11, Gua12b, Str13] and electrochemical resistive switching devices [Yu11, Men12]. Unfortunately, models for at least the former type of devices are still not sufficiently accurate and require further improvement. Such models are typically derived by assuming a particular (typically simple) physical mechanism for resistive switching and electron transport and fitting experimental data to the equations corresponding to this mechanism. For example, in Ref. [Pic09] both dynamic and static equations are fitted assuming modulation of the tunneling barrier width, which is certainly a simplification of the actual physical mechanism and as a result the model does not predict accurately set transition. Because multiple mechanisms are possibly involved in resistive switching [Yan13], such simplification may not always be adequate for accurate models. Additionally, the models are not likely to be general (e.g. judging by the diversity of reported models for the same material system even from the same authors) and future devices may require development of a completely new model from scratch. This is not a marginal issue because as the attempts are made to improve memristive devices the *I-V* behavior may change significantly.

The main contribution of this paper is a development of a general approach for modeling memristive devices. The approach is mainly based on fitting experimental data and hence potentially applicable to broad class of devices featuring smooth non-abrupt switching transition. Using the proposed approach we derive a model for specific titanium dioxide devices. The model is accurate and at the same time simple enough to be suitable for large scale simulations.

## II. General Modeling Approach

The modeling approach is based on several assumptions, which simplify derivation of Equations 1 and 2. The first assumption is to use pulse stress for deriving a dynamic equation. The primary reason is that for a constant voltage pulse with sufficiently short duration $\Delta t$, Equation 1 can be written as

$$\Delta \mathbf{w} \approx G(v, \mathbf{w})\Delta t, \qquad (3)$$

which simplifies derivation of *G(v,w)* by a fitting procedure.

The second assumption, which helps to decouple the derivation of static and dynamic equations, is that practical (nonvolatile) memristive devices have highly nonlinear kinetics [Yan13, Was09] so that it should be possible to measure *I-V* at relatively small biases without causing much disturbance to the memory state. The safe range of voltages depends on the particular type of devices and can be, e.g., determined by performing characterization of the switching kinetics [Pic09, Ali12]. A related assumption is that the memory state is considered to be uniquely characterized by *I-V* measured at small non-disturbing biases (denoted as "read" biases in this paper).

Taking into account the described assumptions, the first step of dynamic equation modeling is the collection of large amounts of data by switching the device with fixed short-duration voltage pulses with different amplitudes and measuring the *I-V* at a non-disturbing bias after each pulse, e.g. similar to the pulse algorithms described in Refs. [Pic09, Ali12]. To simplify the model, it is convenient to use as few state variables as possible, so that only a small number of measurements along non-disturbing *I-V* is required to characterize uniquely the internal state of the device. Ideally, this could be just one state variable, e.g. a measured resistance of the devices $w \equiv R_{V\text{read}}$ at some non-disturbing bias $v_{\text{read}}$. In this ideal case, the objective is to measure the change in state $\Delta R_{V\text{read}}$ for many different combinations of initial state $R_{V\text{read}}$ and pulse amplitude *v*. The actual implementation details of the pulse generation algorithm are not important as long as it will cover all combinations of $R_{V\text{read}}$ and *v*. The next step is to find $G(v, R_{V\text{read}})$ by fitting a surface to $\Delta R_{V\text{read}}$ (*v*, $R_{V\text{read}}$) data. If the $\Delta R_{V\text{read}}$ (*v*, $R_{V\text{read}}$) data are noisy and, e.g., there is a large spread of $\Delta R_{V\text{read}}$ for the same values of $R_{V\text{read}}$ and *v*, then more state variables might be needed, e.g., corresponding to the measured resistance at different non-disturbing biases. In this case separate fitting for each state variable should be performed.

The static equation is modeled by first obtaining *I-V* data from fast non-disturbing sweeps for the device in various initial states, and then fitting the data. For example, in case of single state variable, $F(v, R_{V\text{read}})$ is found by fitting a surface to $i(v, R_{V\text{read}})$ data, where *v* is within a range of voltages used for sweep experiment. Similar to the modeling of dynamic equation, more state variables must be introduced if data are noisy and, e.g., if the device has different *I-V*s for the same $R_{V\text{read}}$. Note that it is important to use as large voltage range as possible without disturbing the state of the device (which can be ensured by checking that the currents for rising and falling directions of voltage sweep overlap), because nonlinear features in static *I-V* are typically prominent at high voltages.

## III. Model for Pt/TiO$_{2-x}$/Pt Memristive Devices

Let us now demonstrate the proposed modeling approach on the example of Pt/TiO$_{2-x}$/Pt memristive devices, whose structure and fabrication methods are described in Ref. [Ali13].[1] For such devices, a single state variable $R_{0.5}$, which represents resistance measured at non-disturbing read bias $v_{\text{read}} = 0.5$V, turns out to be sufficient for good accuracy. Following the proposed approach, the device is switched into different intermediate states by applying a sequence of positive and negative write voltage pulses with $\Delta t = 10$ μs and different amplitudes (Figs. 1a, b). Each write voltage pulse is followed by a read pulse to measure the new device state $R_{0.5} + \Delta R_{0.5}$. The process is repeated for sufficiently large number of different write voltage pulses and initial device states $R_{0.5}$ (with more than 15,000 measurements in total) to gather enough points for fitting procedure. Note that write voltage amplitudes were limited to > -2.5V and < 5V for set and reset transitions, respectively, to avoid breaking the device.

---

[1] Unfortunately, at the moment, we could not test our modelling approach for other types of devices. In general, memristors with repeatable cycle-to-cycle behavior are required for successful modeling and we have only access to high quality titanium dioxide devices developed in our lab.

The resulting 3D plot for resistance change (Fig. 1c) is smooth and features an effective voltage threshold, which justifies using 0.5V bias for non-disturbing read, and convergence to zero for both very high and very low resistances. These features are likely related to Joule-heating-assisted resistive switching [Yan13], e.g. super-linear dependence of temperature increase on applied voltage, redistribution of dissipated power from active region to series resistance upon decrease of resistance [Bor09], and decrease of total dissipated power when resistance increases. Instead of relying on accurate physical model, we use fitting functions $\sinh[v]/(1 + \exp[\chi v + \zeta])$, $R_{0.5}/(1 + \exp[\delta R_{0.5} + \theta])$, and $\exp[\lambda R_{0.5}]$ to mimic these three features, respectively, i.e.

$$\Delta R_{0.5} = \alpha \frac{\sinh[v]}{1+\exp[\chi v+\zeta]} \frac{R_{0.5}}{1+\exp[\delta R_{0.5}+\theta]} \exp[\lambda R_{0.5}] \Delta t, \tag{4}$$

where $\alpha$, $\lambda$, $\delta$, $\theta$ are fitting parameters specific to the direction of switching (inset of Figure 1d). Figure 1d shows the 3D surface based on least square error fitting of Equation 4 to the experimental data. For such fitting, the first term grows super-exponentially with the voltage and emulates an effective switching threshold, the second term is superlinear with $R_{0.5}$ for $R_{0.5} \lesssim 10$ kΩ for set switching and linear with $R_{0.5}$ for both set and reset switching in the remaining range, while the last term introduces an exponential decrease with respect to $R_{0.5}$ in the whole range of resistances for reset switching and for $R_{0.5} \gtrsim 50$ kΩ for set switching. Note that such choice of fitting function is adhoc and primarily motivated by having as few fitting coefficients as possible.

To obtain the static model, the device is first switched into several intermediate states (represented by $R_{0.5}$) by applying positive and negative triangular sweep voltage stimuli (Fig. 2). The static portion of the experimental data $i(v, R_{0.5})$ are then fitted with the following function of $v$ and $R_{0.5}$

$$\log_{10}|i| = g_1 \tanh(1.5 \log_{10}|v|) + \log_{10}|v| + g_2, \tag{5}$$

where $g_1$ and $g_2$ are functions of $R_{0.5}$ (Fig. 3). Similar to a dynamic model derivation, the choice of fitting function for static equation is adhoc and motivated by a tilted tangent hyperbolic shape of the static curves in the log-log scale (Figs. 2b, d). It is worth noting that for either small or large voltages, Equation 5 simplifies to the following linear relation between $i$ and $v$

$$i \approx \begin{cases} 10^{g_2-g_1}v, & |v| \ll 1 \\ 10^{g_2+g_1}v, & |v| \gg 1 \end{cases}. \tag{6}$$

The physical explanation for linear behavior at small voltages is self-evident, while at high voltages it is likely due to the dominant effect of series resistance in memristive devices [Bor09].

Given $G(v,w)$ from Equation 4, the device response to an arbitrary time-varying voltage stimulus can be in principle calculated by solving differential Equation 1. However, another approach is to approximate time-varying voltage stimulus with a sequence of corresponding fixed-duration voltage pulses and use Equation 3 instead. Figure 4 shows simulation results for the full *I-V* sweep using approximated stimuli with the only input parameters to the simulation experiment are initial (measured) state of the device $R_{0.5}$ and applied voltage stimulus. In particular, two cases are simulated and shown on Figure 4. In the first case, only the dynamic equation is utilized (which predicts change in the resistance at 0.5V) and linear current-voltage dependence $i = v/R_{0.5}$ is assumed to get the current at the specific applied voltage. As expected, in this case model is somewhat accurate for small voltages (and hence the static equation may not be needed for modeling) but significantly underestimates current at high voltages. On other hand, the simulated switching *I-V* characteristics are in a good agreement with experimental data in the whole range of voltages if both static and dynamic equations are employed.

## IV. Discussion and Summary

Let us now discuss some limitations and potential improvements for the proposed modeling approach. One reservation concerning using pulse stress (the first assumption of the modeling approach) is that transient effects with slow characteristics times are challenging to model. For example, such transient may be due to a relatively slow heating transient in the device and could be represented by a state variable corresponding to the internal temperature [Pic13, Ber14]. In this case, the device response to a train of pulses will greatly depend on an interpulse delay, even if $\Delta t$ is very small. In the proposed modeling approach slow transients are neglected assuming that there is sufficiently long time $\gg \Delta t$ between applied write and read voltage pulses. Nevertheless, because voltage pulse stimulus is easy to implement in a hardware this simplification is justified for practical applications. Another compelling reason to use pulse train stimulus with large interpulse delay is to eliminate the effect of secondary volatile switching, which is often present in metal oxide devices [Mia11, Ohn11, Cha11, Ber14, Mik14].

It is clear from Figure 1 that there are not much meaningful data for the applied voltages below effective switching threshold and for the device states close to extreme on or off values. In these regions, the changes in $R_{0.5}$ are insignificant and typically too noisy to be used for reliable fitting. To address this issue, the modeling approach can be extended by applying variable duration pulse stress [Pic09, Ali12].

Measured changes in the device state upon application of long-duration voltage pulses can be normalized with respect to $\Delta t$ duration and used reliably for the proposed fitting approach. Additionally, subthreshold behavior might be modeled in adhoc fashion by introducing an additional term in dynamic equation, e.g. based on activation energy of switching [Str09b]. For example, the model predicts that the switching rate for set transition changes by as little as a factor of 900 when applied voltage is reduced from 1V to 0.5V, which is not appropriate for true nonvolatile memory. Multiplying right hand side of Equation 4 by $1/(1+\exp[(v_1 - |v|)/v_2])$, the ratio of switching rates at the same applied voltages is increased to $> 10^{16}$ when using $v_1 = 0.75$ and $v_2 = 0.008$, which is expected retention performance for the considered devices [Ali13]. At the same time, it is easy to check that such extra term will have no effect on the simulation shown on Figure 4.

More generally, the switching model can be approximated with an equivalent circuit shown on Figure 5. It is natural to expect that the maximum resistance of the memristive device is limited by the leakage through the film, which is modeled with resistor in parallel to the active part of the device. Its minimum resistance is determined by a resistance (e.g. corresponding to a filament) connected in series with an active part of the device, which is described by memristive equations. The advantage of such equivalent circuit is that it naturally bounds the device resistance and provides physically plausible switching behavior of the device near extreme on and off states. Additionally, the equivalent circuit can be extended to include secondary volatile switching behavior [Mia11, Ohn11, Cha11, Ber14, Mik14] and electronic noise [Gao12]. Finally, a practical way to include switching variations is to modify parameter $\zeta$ in the model, for example, by adding a zero-mean Gaussian random variable to mimic device-to-device and cycle-to-cycle variations.

In summary, this paper outlines a general approach for deriving memristive equations for resistive switching devices. The approach is purely phenomenological and is based on fitting of the experimental data and hence can be applied to a broad class of memristive devices. The knowledge of the switching mechanisms and electron transport can be helpful for finding the best fitting functions; however, it is not a requirement, which further simplifies modeling approach. The proposed approach is tested on the example of a particular metal oxide device by comparing simulated and experimental *I-V*s for a full sweep. The model shows good accuracy and at the same time, because of explicit form of equations, is computationally inexpensive, which makes it suitable for simulation of large scale memristive circuits.

**Acknowledgments**

This work is partially supported by AFOSR under MURI grant FA9550-12-1-0038, NSF grant CCF-1017579, and Denso Corporation, Japan.

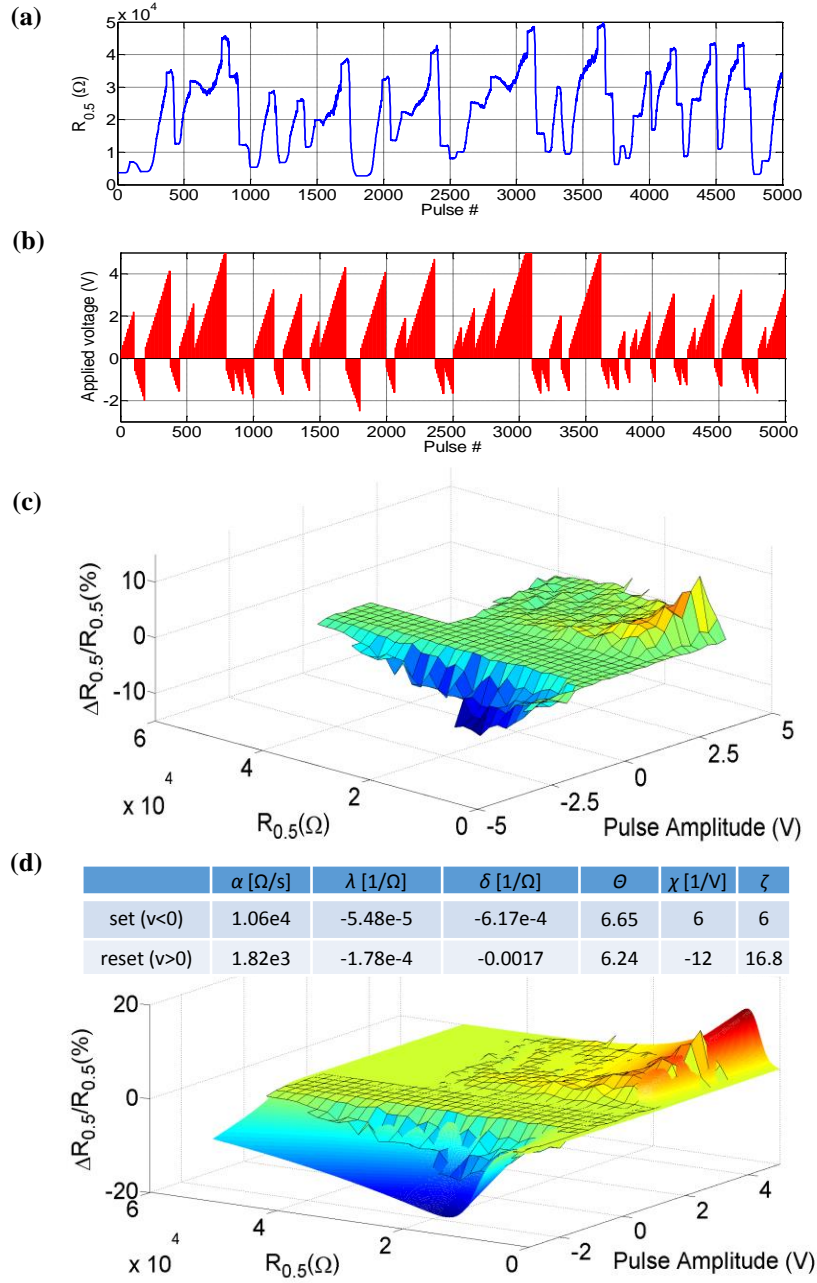

**Figure 1.** (a) Evolution of device resistance measured at 0.5 V as a result of (b) application of sequence of voltage pulses with 10μs write pulse duration and 1s time between pulses. (c) Same as figure 1a shown as a normalized 3D plot (in percent) and (d) fitted surface described by equation 4 with fitting parameters shown in the inset. To reduce the effect of random telegraph noise [Gao12], the resistance measurement is averaged over 20,000 samples taken over 1 ms.

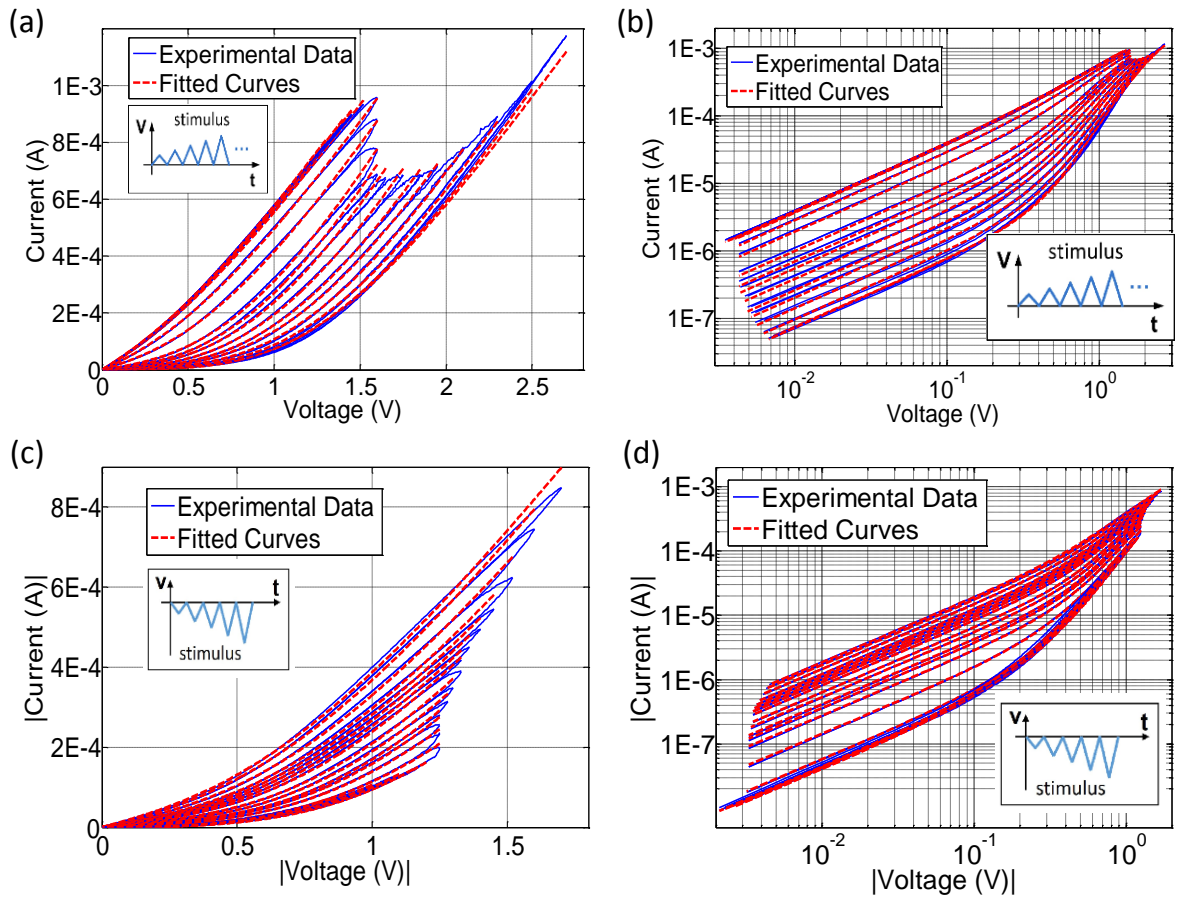

**Figure 2.** Switching *I-V* characteristics and corresponding fitting of static *I-V*s for (a, b) reset and (c, d) set operations. For clarity, figures are shown using (a, c) linear and (b, d) logarithmic scales.

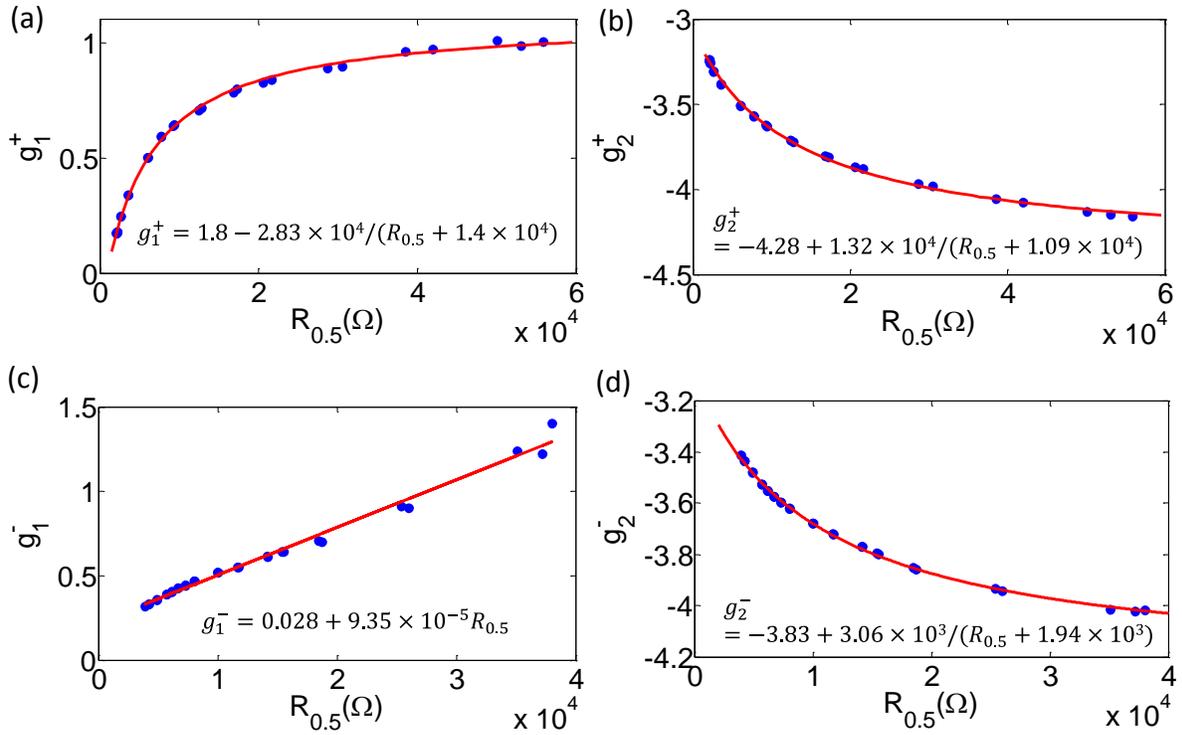

**Figure 3.** Fitting for $g_1$ and $g_2$ functions for (a, b) positive and (c, d) negative voltages. In each panel blue dots are experimental data, while red curve is fitting according to the specific formula.

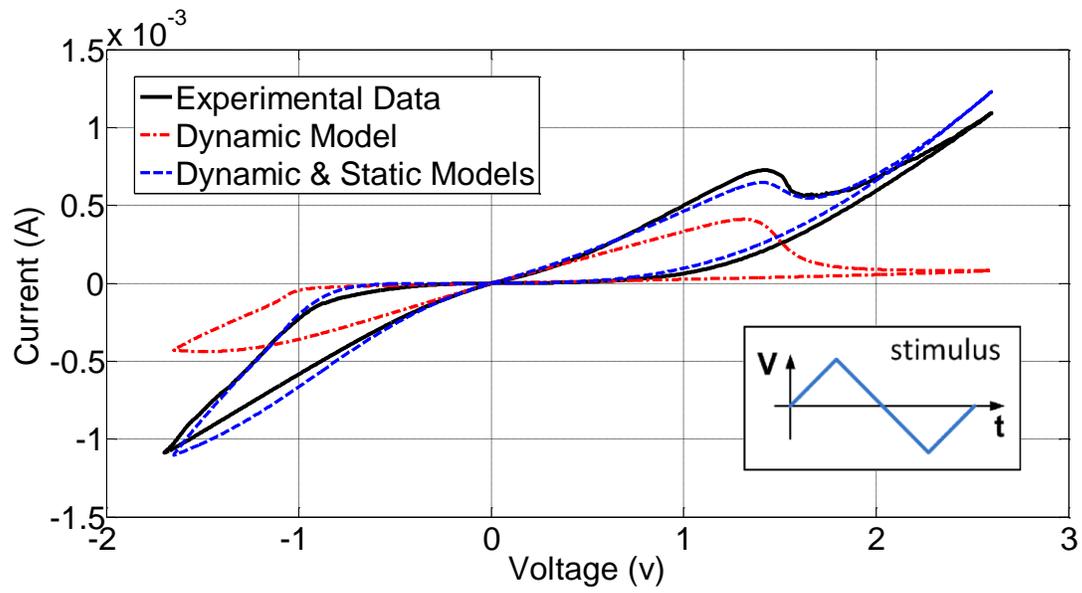

**Figure 4.** Experimental and simulated switching *I-V* characteristics using dynamic equation only and combined dynamic and static equations for triangular voltage sweep.

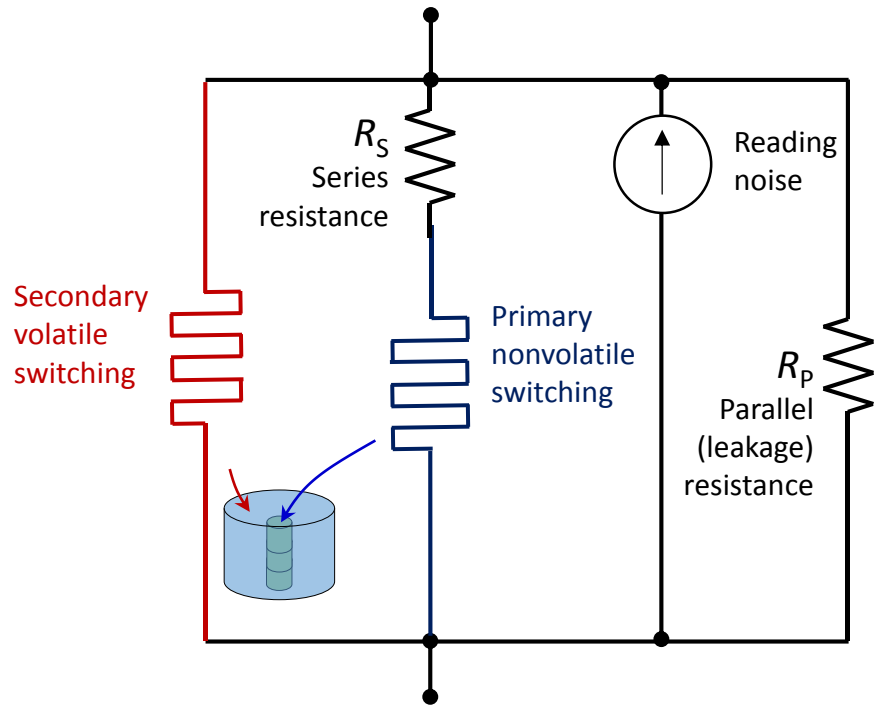

**Figure 5.** Generalized equivalent circuit for modeling metal oxide memristive devices.